# Background Remover - an effective tool for processing noisy microscopy images

A. Kilian*[1], P. Bilski[1], M. Sankowska[1]

1 The Henryk Niewodniczański Institute of Nuclear Physics Polish Academy of Sciences, Kraków, Poland

*Corresponding author: anna.kilian@ifj.edu.pl

**Abstract**

*Background Remover* (*BGR*) is a novel software tool developed as a plugin to the well-known ImageJ program and designed to address the challenges of analysing fluorescent microscopy images characterized by low signal-to-noise ratios and heterogeneous backgrounds. The used algorithm effectively differentiates between signal and noise pixels, preserving the signal while eliminating noise. This functionality enables the analysis of images with objects of varying intensities, allowing for reliable identification even in low signal-to-noise ratio conditions. Furthermore, *BGR* offers the capability to determine the intensity of identified objects, enhancing its utility for researchers in the field. The paper describes the algorithm and the program functioning, as well as the carried out tests of its performance. The program is freely downloadable from the website https://kilianna.github.io/background-remover/

**Introduction**

One of the usual problems in analysing images from a fluorescent microscope is a low signal-to-noise ratio. Exactly such a situation was encountered during work on images obtained from Fluorescent Nuclear Track Detectors (FNTDs) based on lithium fluoride (LiF) crystals.[1] In this type of detector, fluorescence is not emitted by externally added fluorescent probes (dyes), but by crystal lattice defects (colour centres) created by ionizing radiation. In the case of LiF crystals, the most important are $F_2$ colour centres (two anion vacancies with two electrons), which emit red fluorescence peaked at ca. 670 nm, when excited with blue light (around 445 nm). LiF FNTDs are capable of imaging tracks caused by particles passing through detectors e.g. by alpha particles, ion beams from accelerators, neutrons, and even electrons. The details of the LiF FNTD properties and their applications in radiation measurements may be found elsewhere.[2]

The purpose of using FNTDs is not just imaging of nuclear tracks but rather extracting information about the radiation field from a quantitative analysis of track intensity. This intensity is often very low, barely exceeding a background level, which in turn is often not uniform. As the result, the quantitative analysis of such weak fluorescent objects becomes very doubtful[3].

The issues associated with image analysis can be illustrated using the example shown in Figure 1, which features readouts from detectors exposed to cosmic radiation. This image exemplifies certain characteristics that complicate the analysis of routinely captured images. The most significant issue is the heterogeneous background. In particular, certain regions within the background of the images exhibit intensity levels exceeding the signal intensity in other regions. This variability makes it impossible to establish a single, universal threshold for analysis. Furthermore, some images show a significant scatter in the intensity of the recorded tracks, such as the linear tracks with high intensity and smaller circular tracks with noticeably lower intensity in Figure 1. These small circular tracks also demonstrate the issue of low signal-to-noise ratio. In addition, both low-intensity and high-intensity tracks exhibit blurred edges, which further complicates the distinction between signal and noise in the image.

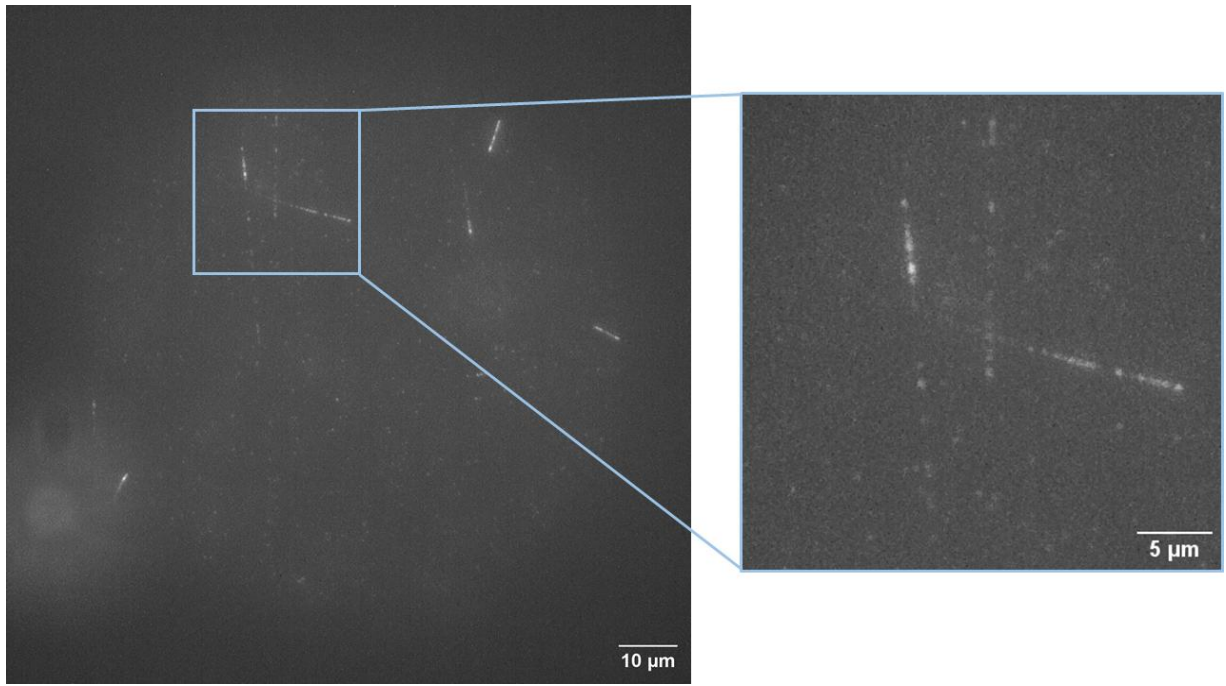

*Fig. 1 Example of an image of fluorescent tracks registered in LiF during standard microscopy readout following exposure to cosmic radiation at the Earth's orbit.*

In addition to the challenges associated with isolating the signal from the noise, a further difficulty is determining the actual intensities of the identified objects. This is further complicated by the difficulty of quantifying the background intensity, which is necessary to determine the intensity of the objects extracted from the image reliably.

There are several numerical methods for subtracting background from images. ImageJ,[7,8,9] the software package most commonly used for the analysis of microscopy fluorescent images, offers some solutions in this regard. However, we have found that the features of the acquired images often lead to unsatisfactory results from publicly available plugins, even those that rely on local threshold calculations. The available background subtraction algorithms change the pixel values of the processed objects (in our case the intensity of the fluorescent tracks), which disturbs the analysis results. Furthermore, although they reduce the image background significantly, they do not remove it entirely, which would be highly desirable for the quantitative analysis of fluorescent tracks.

Over the years, we have tested several plugins and functions for background separation in ImageJ. For instance, Figure 2 compares the best outcomes of Figure 1 transformation achieved using ImageJ's built-in function *Subtract Background* (Fig. 2a) and the Mosaic Suite's[4] *Background Subtractor* (Fig. 2b). In both cases, there was a significant reduction in the background noise, however, some noise persisted, making it difficult to distinguish the signal, particularly in the areas with a low signal-to-noise ratio.

Considering the issues mentioned above, we decided to develop a dedicated method for this purpose and implement it as an ImageJ plugin. The primary requirement for the software was that, after the background subtraction, any changes in the intensity of the fluorescent tracks, if any, would be negligible.

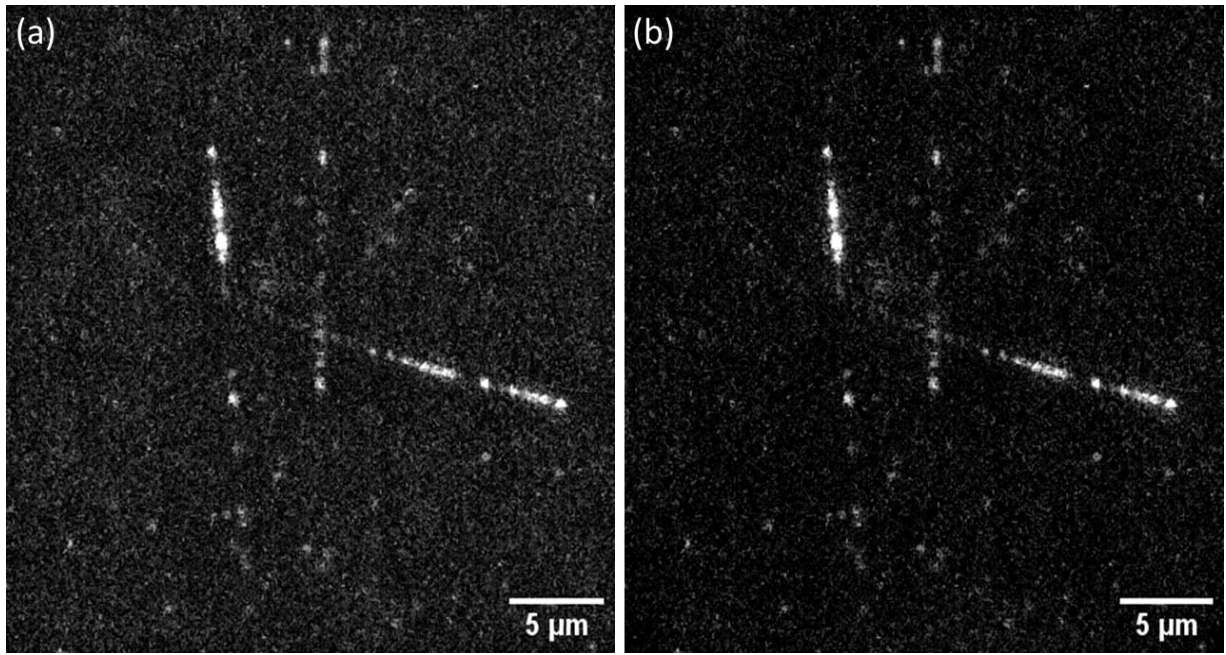

*Fig. 2. Comparison of the performance of two standard ImageJ background subtraction procedures: (a) - ImageJ's built-in function "Subtract Background", (b) - Mosaic Suite's "Background Subtractor". The processed image is the cutout of Figure 1.*

The developed method is based on recognizing objects considered as tracks and removing all reminders from the image. The method was originally developed for analysing tracks having the form of small, circular-like objects, but it was then found that it works equally well for larger and linear-shaped tracks and therefore seems to be quite universal.

This paper describes the algorithm we developed and its implementation as an easy-to-use ImageJ plugin named *Background Remover*. The results of tests of its performance in processing FNTD images are also reported.

**Methodology**

*Sample preparation and microscopy observations*

FNTDs were prepared from LiF single crystals produced at the Institute of Nuclear Physics in Kraków using the Czochralski method. The starting material for samples analysed in our research was LiF powder. The grown crystals were later sliced into small samples using diamond saws, with each sample measuring a standard size of 4 × 4 × 1 mm. For more details on standard sample preparation, please refer to our previous works[1,5].

Microscopy examinations were conducted using a Nikon Eclipse Ni wide-field microscope (Tokyo, Japan) paired with a CCD DS-Qi2 camera. The illumination was provided by a pE-100 system featuring 440 nm LEDs (CoolLED), along with a band-pass filter ET445/30. For the emission light, a long-pass filter ET570lp was utilized. All images were captured with a 100× TU Plan ELWD objective lens (NA 0.80). The microscope is equipped with an adjustable diaphragm that restricts the field of view to a nearly

circular area with a diameter of 90 μm, corresponding to an observed area of approximately 6900 μm².[1,6]

The registered images were stored in the .ND2 file format, which is used for saving images and metadata generated by Nikon microscope cameras. The recorded images were 14-bit grayscale with a resolution of 13.7 pixels per micron. Most of the images captured were in the form of stacks, consisting of images registered with the focus set at different focal depths within the crystal.

In this study, the intensity of a pixel is represented by a number ranging from 0 to 16383, which corresponds to 16383 shades of gray in a 14-bit image, which is the standard resolution of our microscope output files. This intensity value reflects pixel brightness, where an intensity of zero indicates the colour black, and the maximum intensity represents the colour white. The area of an object in an image is measured in pixels, indicating the total number of pixels that constitute the object. The Signal-to-Noise Ratio (SNR) is defined as the ratio of the intensity of the generated points to the average background intensity.

*General plugin information*

The new software tool, *Background Remover* (*BGR*), was created as a plugin for the well-known, free, open-source image processing software ImageJ[7,8,9] and its distribution Fiji.[10] The Fiji package comes with various useful plugins that aid in processing and analysing scientific images. As an open-source project under the GNU General Public License, Fiji allows users to contribute and enhance its functionality by adding their own plugins. The *BGR* plugin is hosted in the Git version control repository (https://github.com/kilianna/background-remover/) and offers comprehensive documentation. Both ImageJ/Fiji and *Background Remover* are Java-based and can be run on different platforms, including Windows, Linux, and Mac OS X. In this paper, we will present its main functionality, operation principle, and use examples. The download link and further information are included in the plugin website (https://kilianna.github.io/background-remover/). The supplementary materials include a manual and a video that demonstrates the plugin's *BGR* operation on a sample file.

*Background Remover* can process 16-bit grayscale images. The plugin can be applied to a single image or a stack of images of the same size. *Background Remover* is suitable for images with a dark background and bright objects. At the output, we can receive a processed single image or a stack, also in a 16-bit version.

*A general idea of the method*

The main idea of the *Background Remover* is to determine for each pixel in the image whether it belongs to a relevant object (signal pixels) or to noise (noise pixels), and then to set all noise pixels to zero. To make this determination, the average intensity of the surrounding pixels at different distances from the examined pixel is calculated. In general, if the average intensity of the pixels in the immediate vicinity of the examined pixel decreases with increasing distance from the examined pixel, then it is classified as a signal pixel. Otherwise, it is assumed to be a noise pixel. This general operating principle is illustrated in Figure 3. Note that the classification of a pixel as part of the signal or noise is influenced by its local environment rather than the entire image. This approach helps to keep the result unaffected by variations in the background. At the same time, based on the local values, we are able to apply a consistent criterion for pixel discrimination across the entire image.

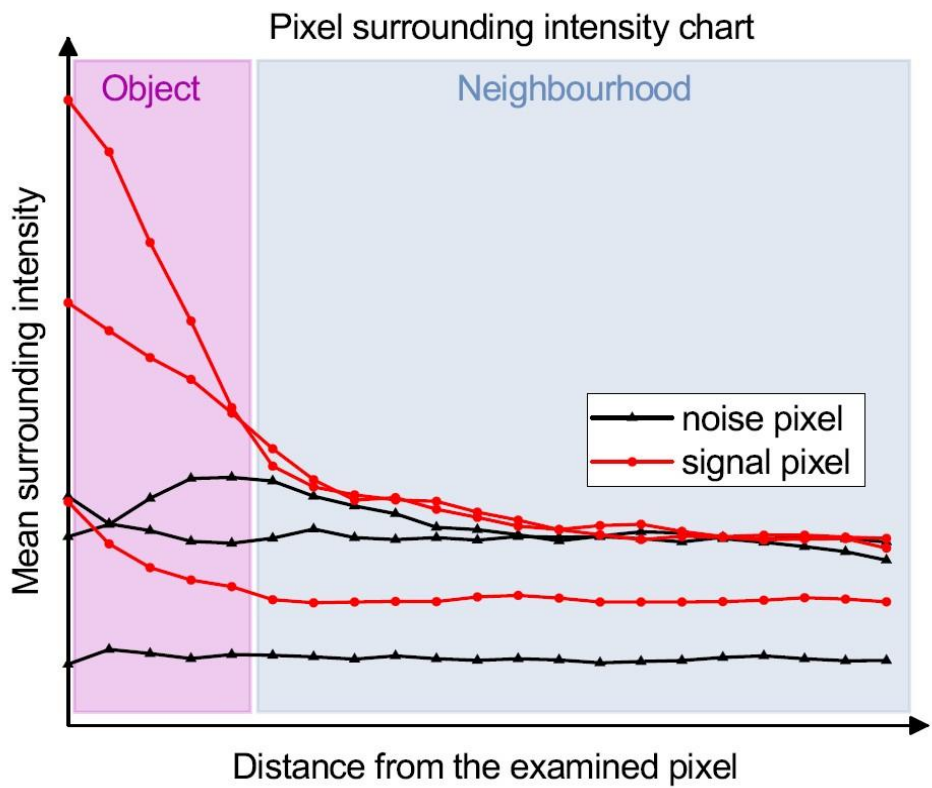

*Fig. 3 Example curves illustrating the dependence of the average intensity of the pixel's surroundings on the distance from this pixel. The red ones were classified as signal pixel curves and the black ones as noise pixel curves.*

As can be seen in Figure 3, our approach allows us to assign a pixel as a signal pixel even if the absolute intensity of its surrounding is lower than the absolute intensity of the noise pixel in different parts of the studied image.

The plugin has a distinct feature that allows it to estimate the true intensity of recorded signal pixels. This is achieved by subtracting the estimated local background of the objects. The plugin treats adjacent pixels as a single object and calculates the average background intensity surrounding these identified objects. A user can adjust the parameters for the location and size of the local background, which helps to minimize the effect of blurred edges of objects.

*Implementation of the method*

The *Background Remover* algorithm consists of two main components: the pixel surrounding intensity calculation stage and the discrimination stage (see Fig. 4). In the first stage, the algorithm calculates the average signal intensities in the vicinity of each pixel within a specific radius. This process generates a series of graphs, so-called "Pixel surrounding intensity charts" (see Fig. 3), which plot signal intensity against the distance from the pixel, with one chart created for each pixel. The second stage, which relies on the results from the first part, aims to identify which pixels are classified as "signal pixels" and which are considered "background/noise pixels."

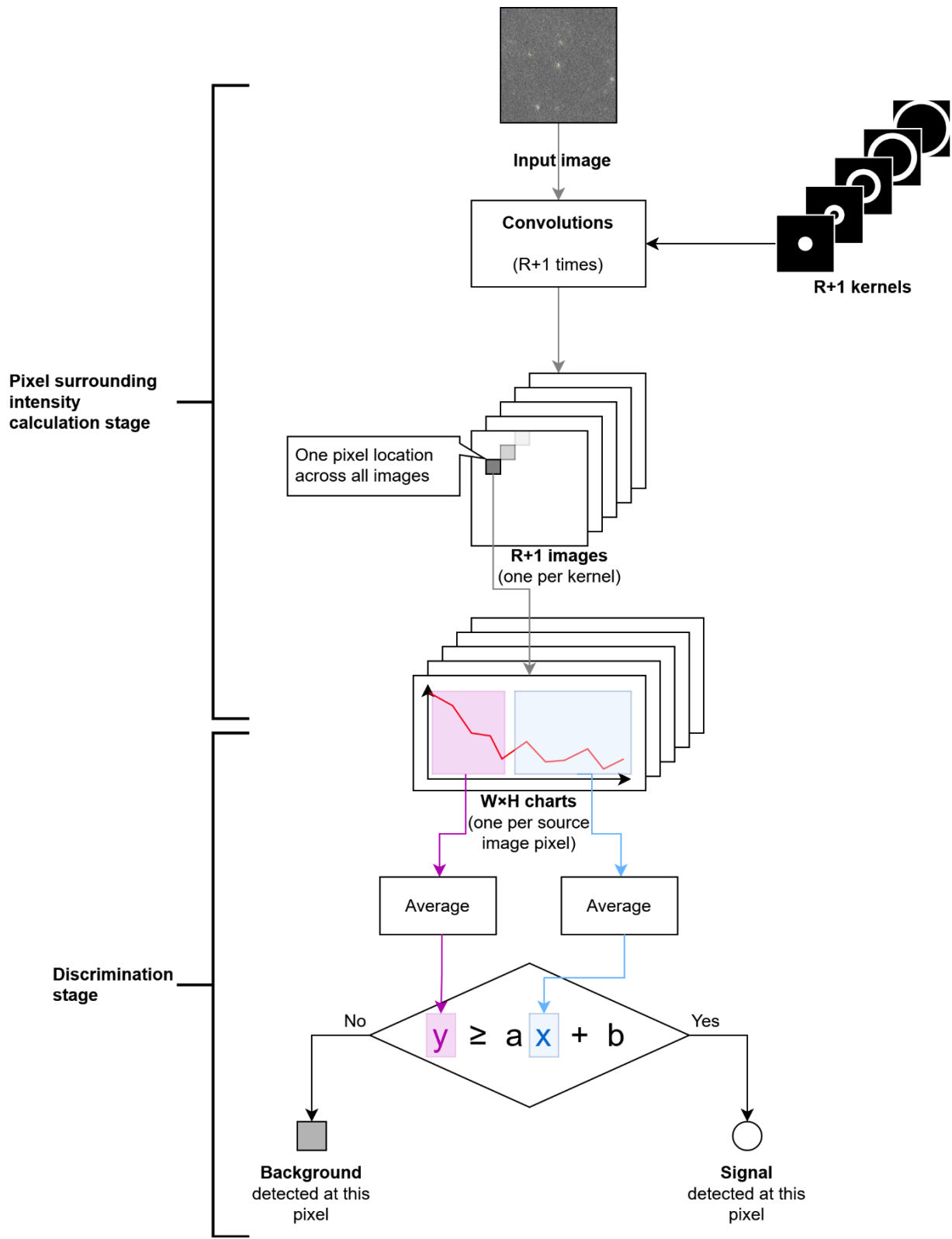

*Fig. 4 Flowchart of the BGR algorithm.*

The first module focuses on convolving the input image with specific kernels that correspond to rings with progressively increasing radii. Image convolution involves applying a small matrix, known as a kernel, to the image by sliding it over the pixels and processing each position based on the values of neighbouring pixels. At the edges of the image, calculations are done using a reduced number of surrounding points. The number of convolutions (R+1) performed for each pixel is determined by the user and depends on the size (2R+1) of the selected image scanning window. More detailed information about the transformations performed and the kernels used can be found in the Appendix.

As a result of convolution, R+1 images are generated. The intensity values of pixels that occupy the same location across these images are combined. This process yields a "Pixel surrounding intensity chart". The number of pixels in the graph is equal to the number of pixels in the original input image: W x H (where W and H are the width and height of the input image). The X-axis of the graph represents the different radii from a specific pixel, while the Y-axis indicates the average intensity of pixels over each given radius (see Fig. 3).

The discrimination stage takes the pixel surrounding intensity graph as input and determines whether the associated pixel should be classified as background or signal. The process begins by dividing the graph into two regions: one for pixels that are in a closer surrounding area (Object area) and another for pixels that are farther away (Neighbourhoods area). The boundary between these two regions is set manually based on the size of the objects being analysed.

Next, the average intensities of each region are calculated. The result for the closer region is denoted as y, while the result for the farther region is denoted as x. Discrimination is performed using the linear equation: $y = a \cdot x + b$, where a and b are manually selected parameters. The plugin assists with the adjustment of a and b, making it easy to choose appropriate values for specific types of images.

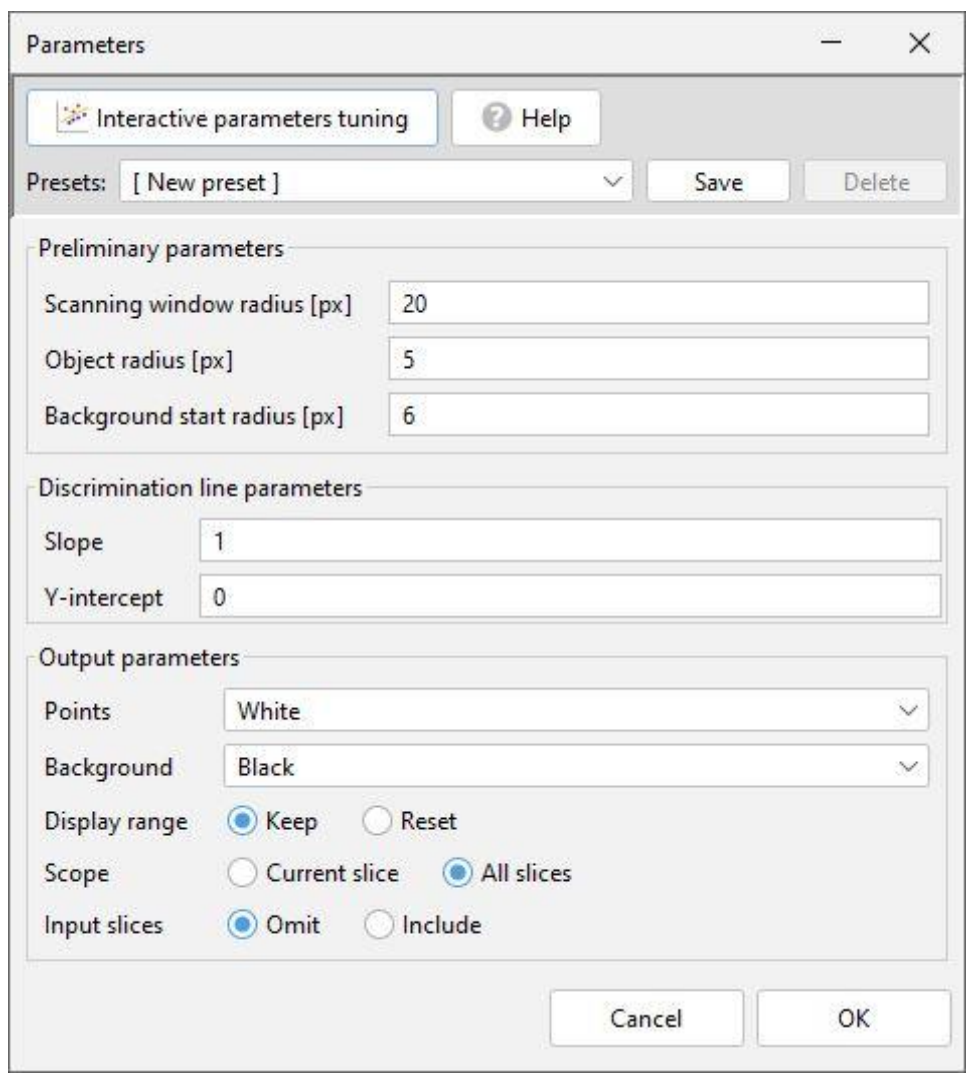

Fig. 5 Main window of the plugin.

*Program description*

The *BGR* plugin was designed to be straightforward, making its operation simple and minimizing the number of parameters. At the same time, it allows users to select various options to optimize performance for a wide range of different images. A user-friendly graphical interface of the program is shown in Figure 5.

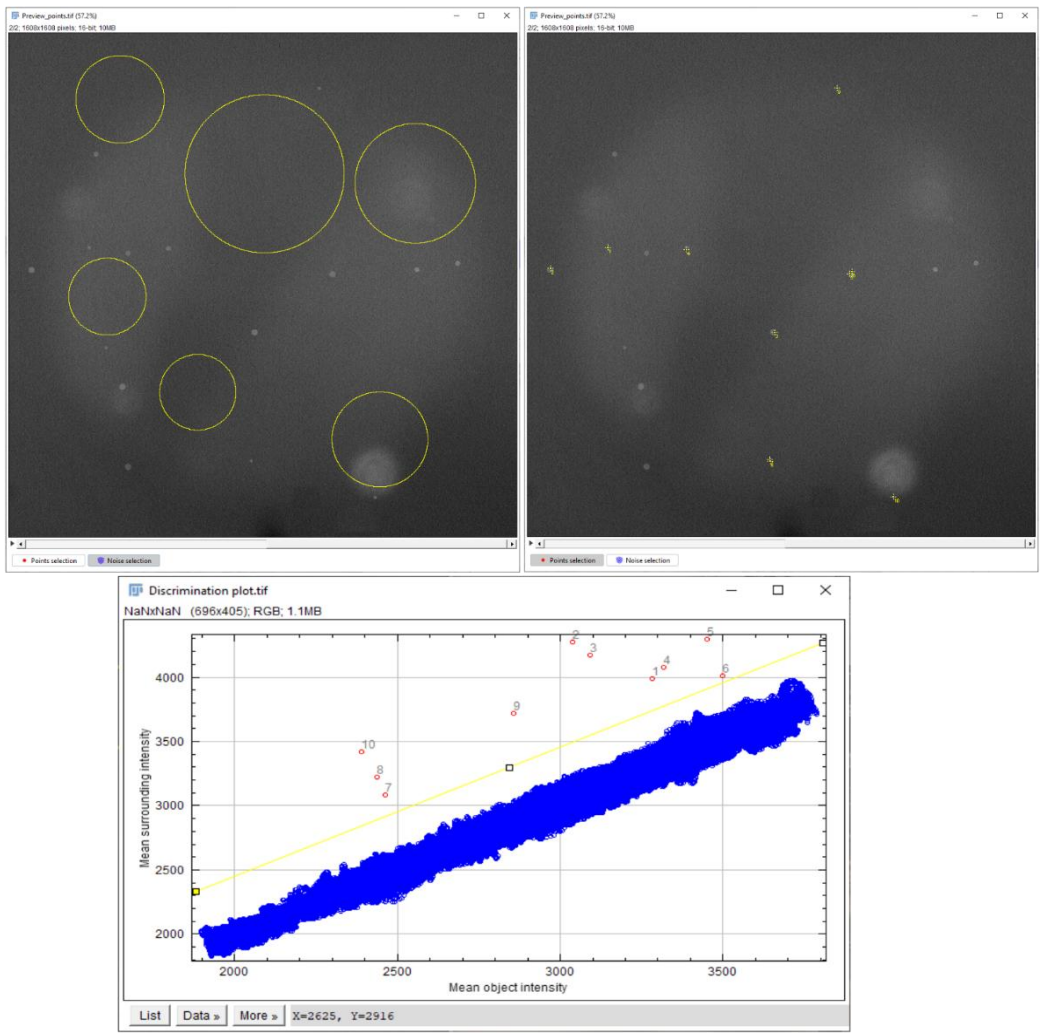

*Fig. 6 Screenshots from the manual version of the BGR plugin. Top left - manually marking background pixels, top right - manually marking object pixels, bottom – discrimination plot.*

The program offers two modes: automatic and manual. In automatic mode, users can enter predefined parameters in individual fields or load a previously saved set of parameters. In manual mode, users can optimize all parameters using additional specialized tools designed to assist in the optimization process. Figure 6 displays sample screenshots from manual mode. The top row features a tool that enables the user to mark areas of the analysed image designated as background (shown in the left image) and pixels identified as belonging to objects (shown in the right-hand image). The bottom image displays the pixels selected by the user as background and as objects on a graph. The graph shows the relationship between the intensity of each pixel and the intensity of its surrounding. This visualization enables the user to see the discrimination curve and optimize its parameters. Additionally, the plugin can suggest a discrimination curve based on the areas specified by the user. In manual mode, a preview of the expected results is also available.

The final output image produced by the plugin may contain different types of information. One of the simplest options is to generate a black and white image that indicates the locations of the detected objects. This is a helpful approach for creating masks that can be utilized for further image processing, first of all for just counting objects. The output image may also contain the original intensity values of the identified objects. Another interesting option is to produce an output image that displays the probability with which a given pixel is classified as a signal pixel (Degree of matching option). This feature is achieved by illustrating the difference between the left and right sides of the discrimination line equation.

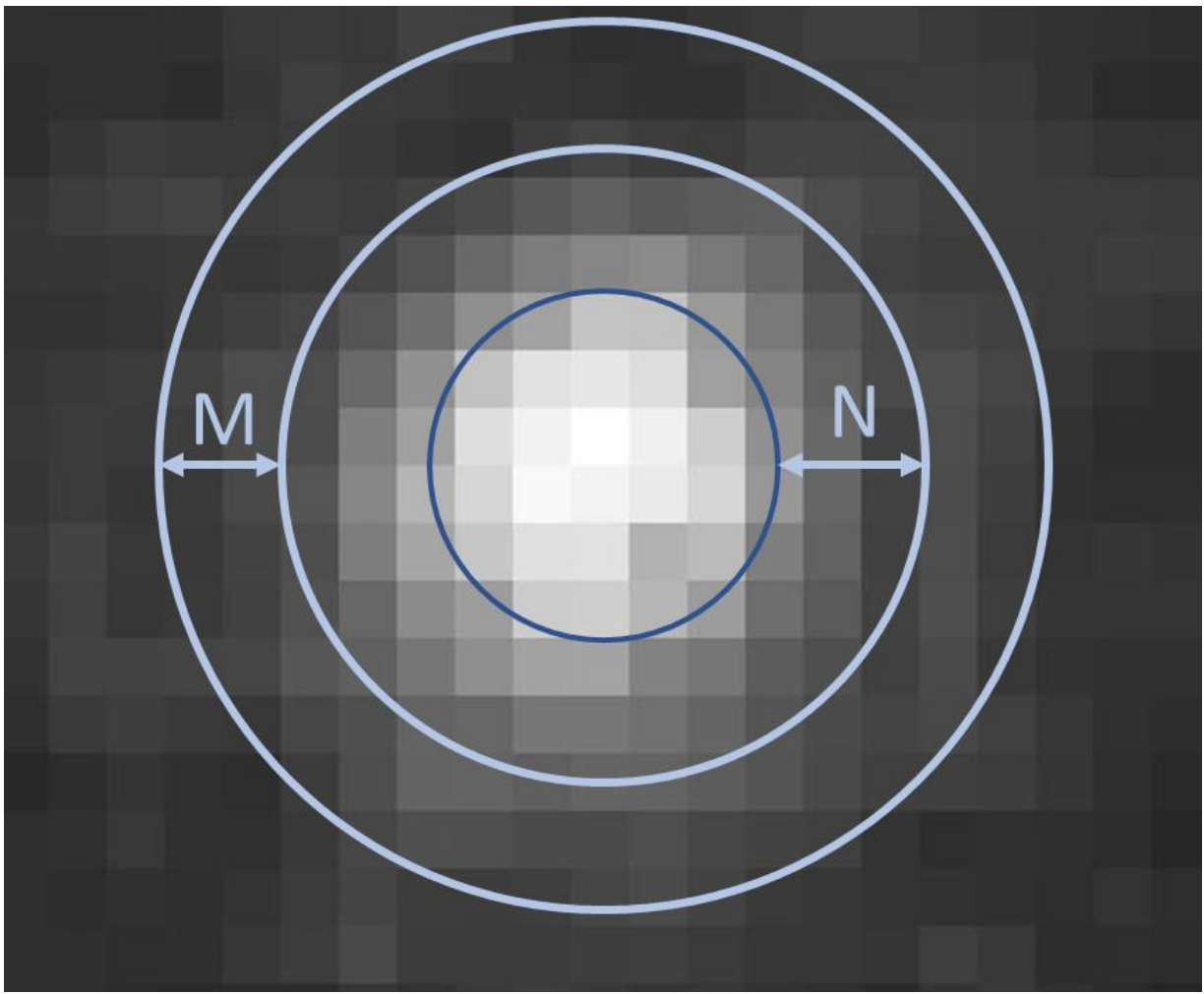

*Fig. 7 Envelopes used for background estimation.*

However, from our perspective, the most useful option is the calculation of the Net Signal. The net signal generation begins with grouping pixels that are identified as signal sources. Each group consisting of adjacent signal pixels is recognized as an object. The next step involves finding the envelope around each identified object. This envelope is N pixels thick (see Fig. 7), determined by the "skip pixels" parameter. Pixels included in this envelope should not be considered for quantifying background, as they are too close to the objects and may introduce bias. The second envelope is M pixels thick, based on the "take pixels" parameter. Pixels within this envelope are regarded as background for the specified object. The output object pixel intensity is calculated by subtracting the average, mode, or median (to be chosen by the user) of the pixels within the second envelope from the original object pixel intensity.

*General information about tests*

In order to quantitatively verify the performance of the *Background Remover*, the plugin was subjected to several tests. The tests have been designed to reflect as closely as possible the actual conditions of use of the FNTDs, while allowing control over selected parameters of the objects studied. Unless otherwise stated in the text all tests were performed on images consisting of a chosen background image originating from a real measurement of an unirradiated FNTDs, on which artificially generated objects were superimposed. These objects were created as non-overlapping circles with the user-defined radius and intensities. Their position was selected randomly.

Figure 8 illustrates the process of creating the images used in the experiments. First, an image of the actual background from routine measurements on a lithium fluoride crystal sample was selected Figure 8a. Next, 15 points with varying intensities and radii were randomly generated Figure 8b and placed within the image. Finally, Figure 8c represents the combined result of images Figure 8b and Figure 8c.

The combination of background and generated points can be approached in two different ways. In the first variant, the intensities of the generated circles are simply added to the original intensities of the background pixels. This results in objects with non-uniform intensity, which varies depending on the background where the object was created. In the second variant, the intensity of the objects in the output image is determined by adding a user-defined intensity to the average intensity of the background pixels at the point of creation. The resulting objects have uniform intensity over their entire area. Both of these approaches were used in this study for different purposes.

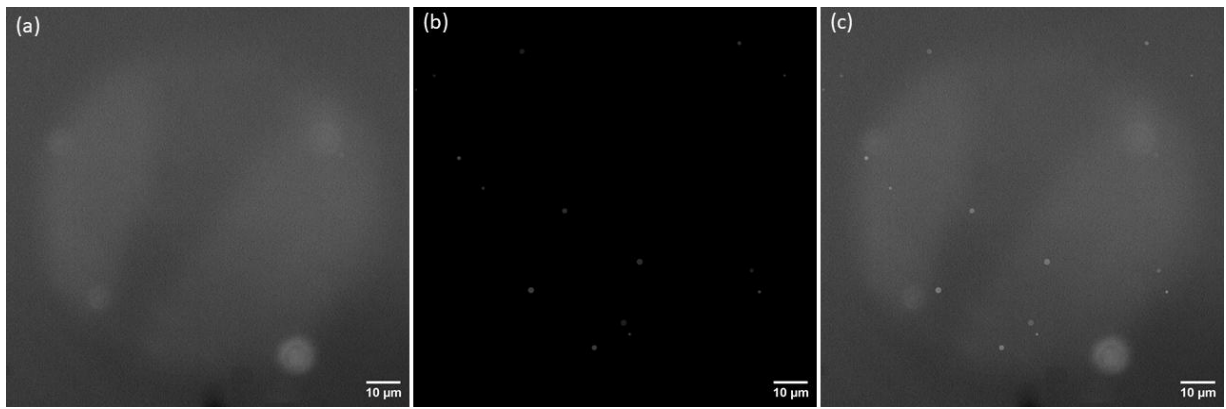

*Fig. 8 Example of creating an image for analysis: (a) background from real microscopy measurements, (b) randomly generated objects, (c) resulting combination of images (a) and (b).*

**Results and discussion**

*Examples of plugin performance*

Below are presented some examples of using *BGR* on various images captured during routine measurements. Figure 9 presents the performance of our plugin on the cutout of Figure 1. This image shows tracks of different types of charged particles from cosmic radiation, recorded during the six-month period that lithium fluoride crystals spent on the International Space Station (ISS).[11] The recorded particles vary in type and energy, as well as in their angles of incidence. Consequently, the traces captured in the image appear as both straight lines and points. Additionally, the intensity of the recorded tracks varies significantly. It is evident that both high-intensity longitudinal traces and lower-intensity point traces were successfully extracted. Additionally, the background was completely eliminated, which was not achievable with the previously tested built-in plugins (cf. Fig. 2).

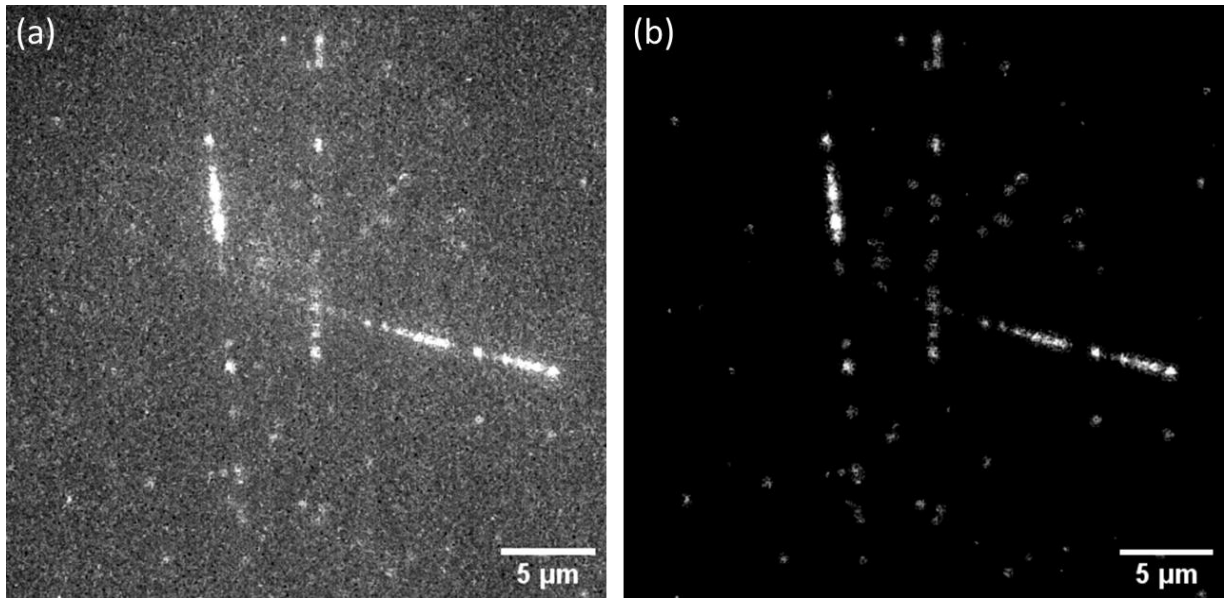

*Fig. 9 Example of performance of BGR plugin on the cutout of Figure 1. The original image (a) and transformed with BGR plugin (b) image.*

Figure 10 displays an image captured after LiF crystal irradiation with an accelerator beam of O-16 ions.[12] The recorded tracks appear as small round objects because the ion beam was directed perpendicularly to the crystal surface. Figure 10a shows the original image, while Figure 10b presents the same image processed using *Background Remover*. Figure 10b clearly demonstrates that all objects were accurately identified while unnecessary background elements were removed.

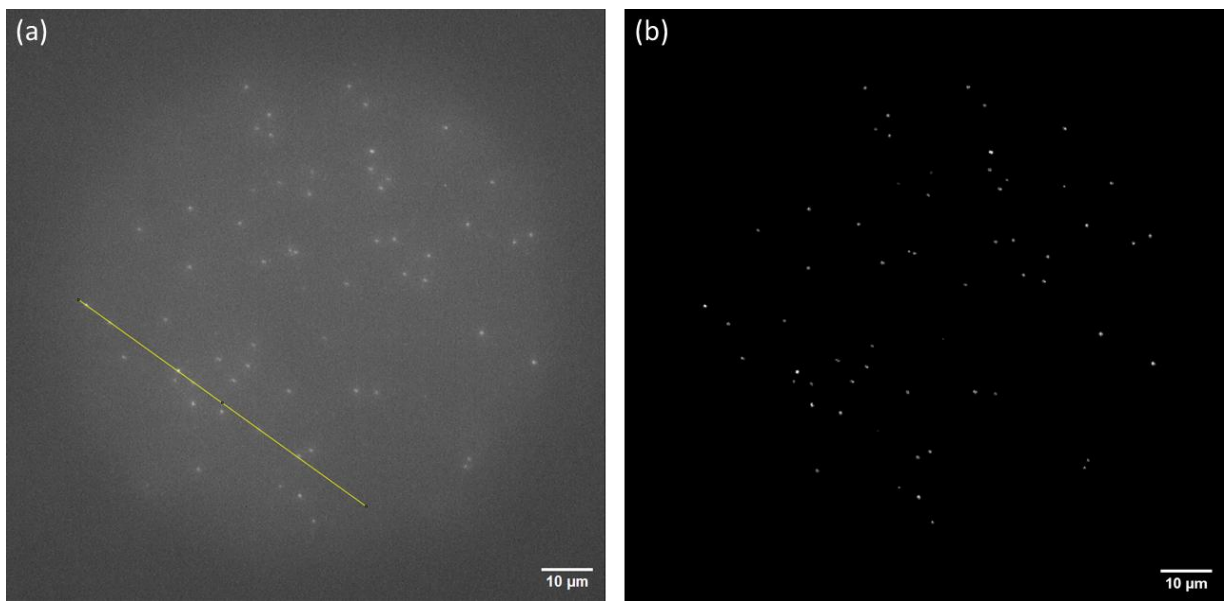

*Fig. 10 Example of performance of BGR plugin on the image registered after exposure to O-16 ions: (a) - original image, (b) image transformed with BGR plugin. The yellow line indicates a line along which the intensity profile was plotted (see Fig. 11).*

The same original image was also processed using the Mosaic Suite's *Background Subtractor* plugin and the *Subtract Background* function. The yellow line marked in Figure 10a indicates the area from which the plot profile presented in Figure 11 was generated.

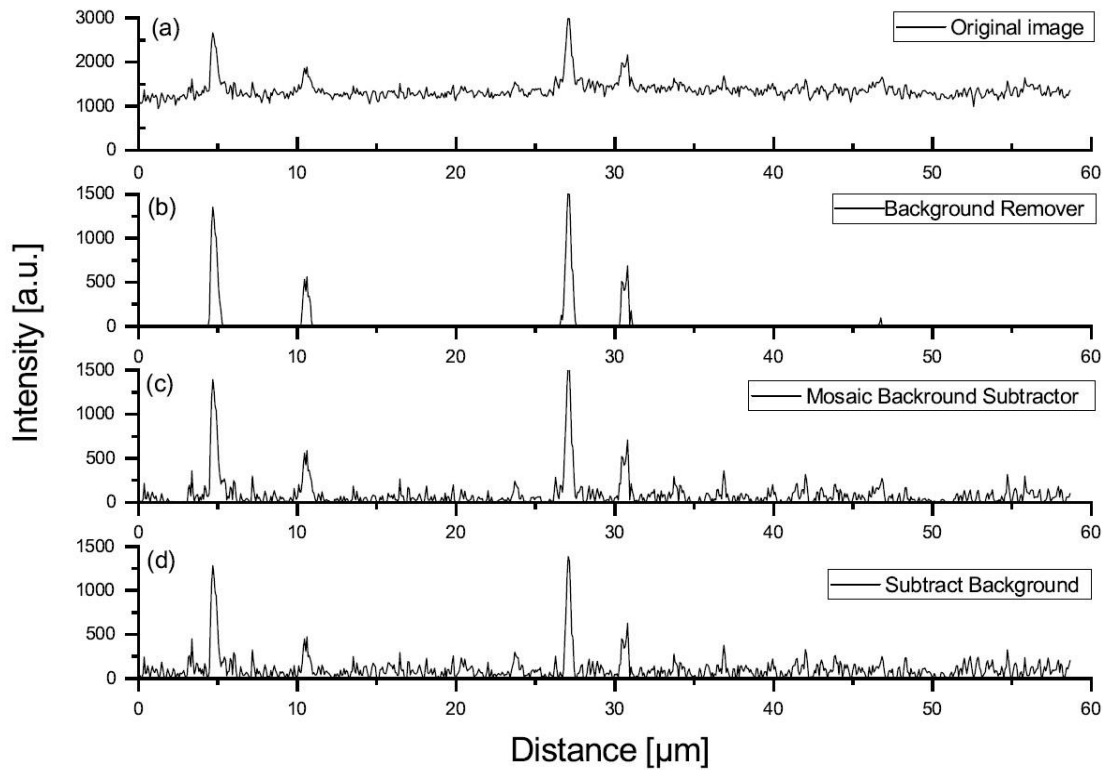

*Fig. 11 Comparison of the intensity profiles of the original image and images obtained after processing with different plugins: (a) original image, (b) Background Remover, (c) Mosaic Background Subtractor and (d) ImageJ built-in function Subtract Background. The pixel intensity values were taken along the line marked on Fig. 10.*

The graph in Figure 11 compares the intensity changes along the profile for the original image and the images processed with plugins. In all three cases, there is a noticeable reduction in the background level. However, it was completely eliminated only in the case of *BGR*. The complete removal of the background is especially crucial for images with a low signal-to-noise ratio. In such challenging cases the *BGR* plugin reveals its full capabilities. An example of such an image is presented in Figure 12a. This image displays the tracks recorded after exposing FNTDs to protons, with the proton beam directed parallel to the exposed crystal.[12] The SNR is so low that distinguishing objects from the background is hardly possible with the naked eye. However, by using the *BGR* method, we were able to effectively extract the signal from this noisy image. The yellow rectangle marked in Figure 12b indicates the area from which the plot profile presented in Figure 13 was generated. The above conclusions are also confirmed by the graph in Figure 13. The plot displays a "column average plot", where the x-axis represents the horizontal distance through the selection and the y-axis the vertically averaged pixel intensity.

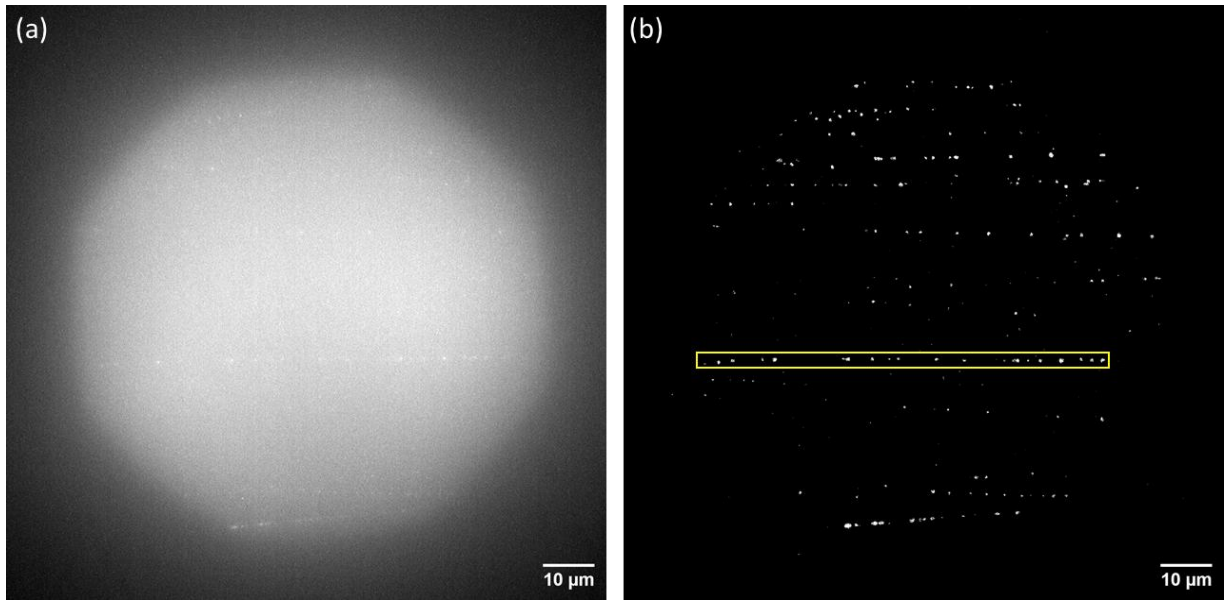

Fig. 12 Example of performance of BGR plugin on image registered after exposition to protons: (a) - original image, (b) image transformed with BGR plugin. The yellow rectangle indicates a region used for plotting the intensity profile (see Fig. 13).

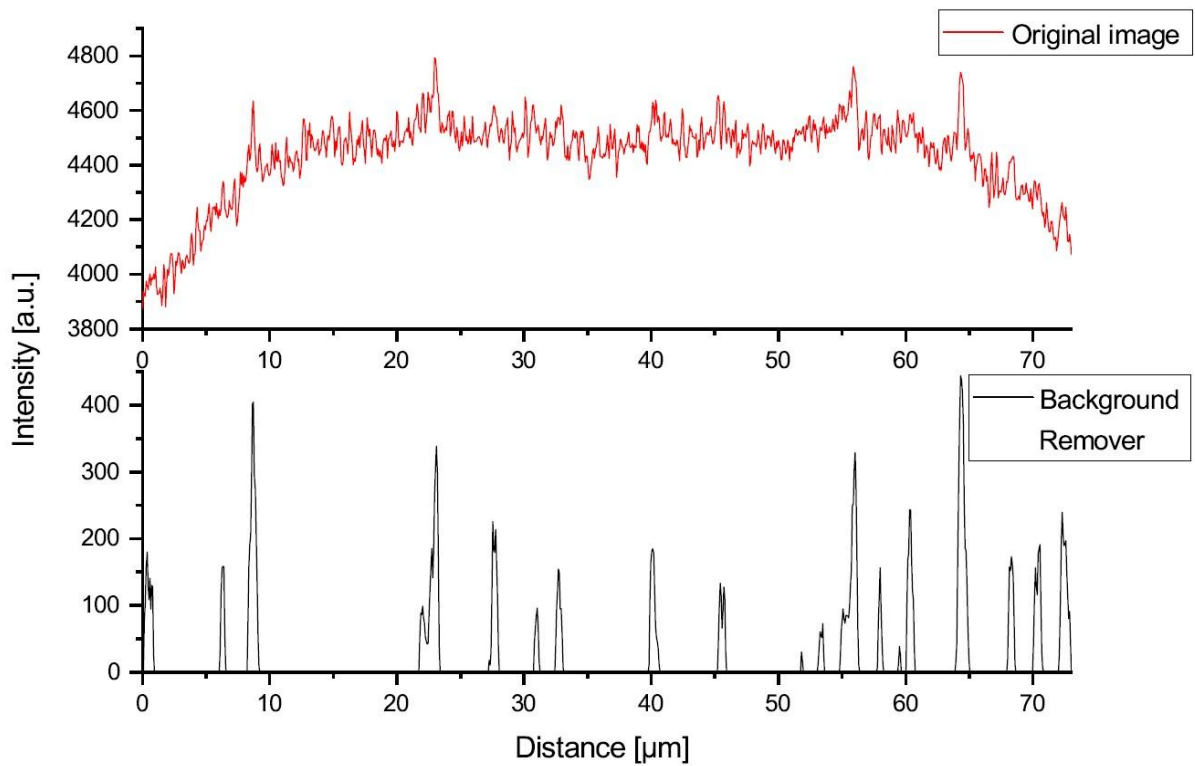

Fig. 13 Comparison of the intensity profiles of the original image and image obtained after processing with Background Remover. The pixel intensity values were taken along the rectangle marked on Fig. 12.

*Test of effectiveness of object detection*

The first quantitative test aimed to evaluate how effectively the plugin can detect objects in an image with a highly diverse background. The background selected for the study included areas with varying intensity levels. The background intensity levels ranged from 0 to 4800, with an average of about 2700. Circle-shaped objects, with radii of either 5 or 10 pixels, were randomly generated across the background image. The intensity values of the individual pixels within the generated objects ranged from 300 to 1000 (i.e. such values were added to the original intensities of the background pixels), which resulted in SNR of 0.11 to 0.37. A stack of a total of 100 images was generated, each containing 20 objects, resulting in 2000 objects with specific radius-intensity parameter pairs. Figure 14 presents two examples of objects generated on different backgrounds: a) objects with a 5 px radius and an intensity addition of 1000 (SNR 0.37) and b) objects with a 10 px radius and an intensity addition of 500 (SNR 0.18).

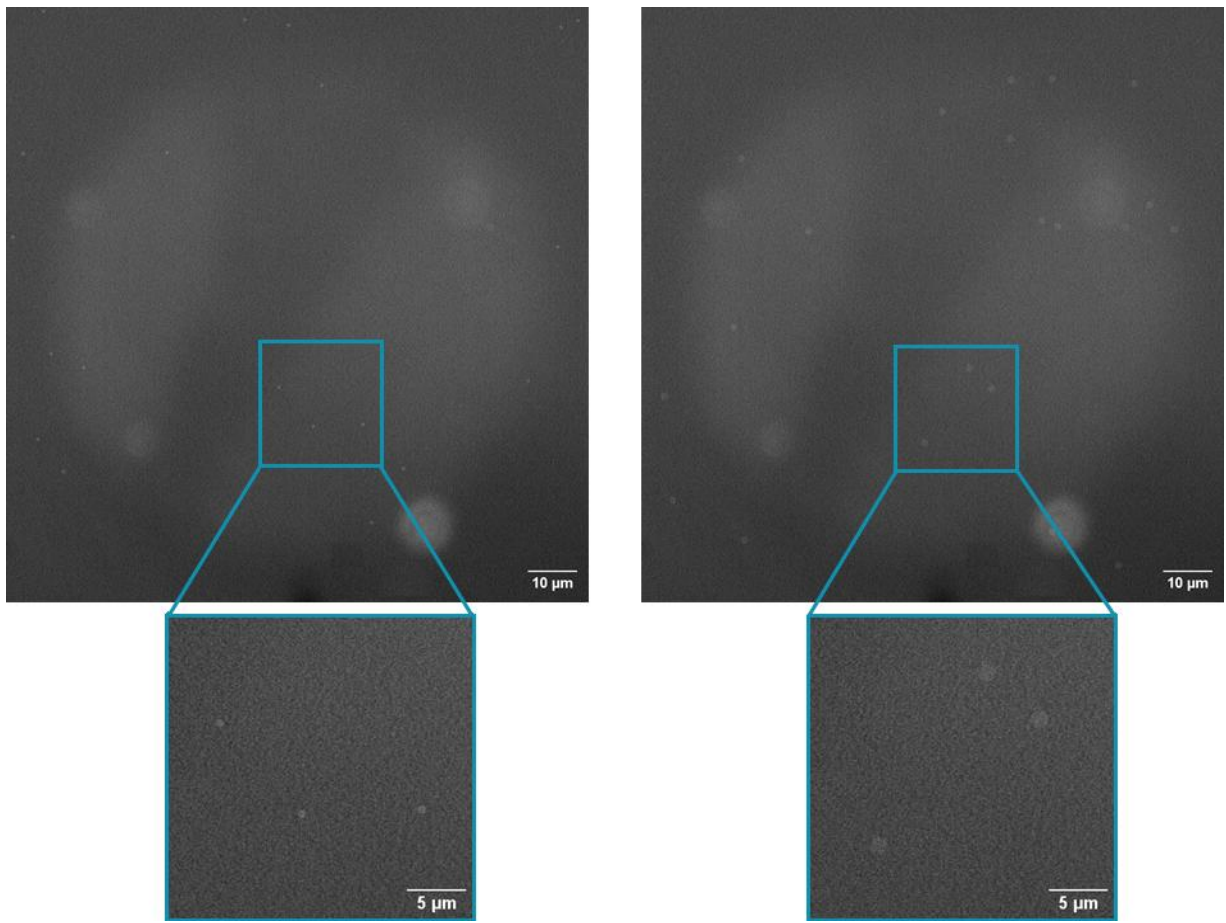

*Fig. 14 Examples of objects generated on varied backgrounds: a) objects with a 5px radius and an added intensity of 1000; b) objects with a 10px radius and an added intensity of 500.*

For each stack obtained, the manual option of *Background Remover* was used to determine the most optimal parameters, which enabled the detection of generated objects in all 1000 images. To facilitate further analysis, the results were exported as white objects on a black background. Subsequently, the resulting series of images were analysed using a built-in function in ImageJ called *Analyze Particles*. The detected objects were not only counted, but also their area was determined. It is important to highlight that the generated output image did not require any additional pre-processing for calculating the number of objects. Furthermore, there was no need to impose any additional restrictions in the

*Analyze Particles* function, such as defining a minimum size for detected objects or considering objects' circularity.

Figure 15 presents the obtained dependence between the number of detected objects and SNR for objects with radii of 10 px and 5 px. The results indicate that for an SNR greater than 0.2, all generated objects with a radius of 10 px and 5 px were detected accurately. However, below the SNR threshold of 0.2, the number of detected objects with a radius of 5 px begins to decrease rapidly, falling to less than 10% detection at an SNR of 0.11. For objects with a radius of 10 px, 100% detection is achieved up to an SNR of 0.17. Beyond this point, there is a slight over-detection of objects, followed by a decrease in the detection percentage, similar to what was observed with the smaller objects.

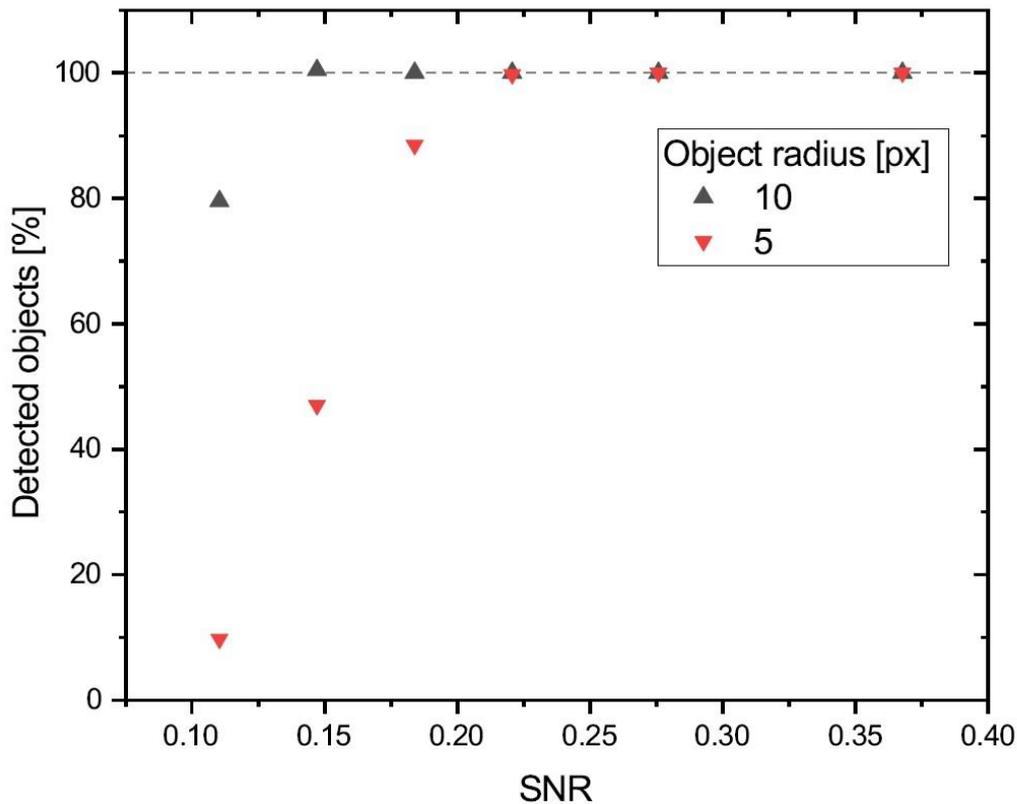

*Fig. 15 Dependence between the number of detected objects and SNR for objects with radii of 10 px and 5 px.*

This slight increase in the number of detected objects occurs because, at lower intensities, relatively large objects can be divided into two smaller ones, resulting in one object being counted twice.

A side effect of the implemented algorithm is that for low values of SNR the area of detected object is reduced. This is illustrated in Figure 16, which presents areas of detected objects with a nominal radius of 10 pixels across three different intensity levels. However, it should be noted, that the discrimination line parameters were chosen primarily to optimize the accurate representation of the number of detected objects, rather than their precise area. While it is possible to enhance the representation of the area of the detected objects, doing so would likely lead to the detection of false objects, which would then need to be filtered out using additional methods, such as implementing size-related discrimination criteria. For our applications correct determination of the number of objects has clear priority, as object areas are of lesser importance.

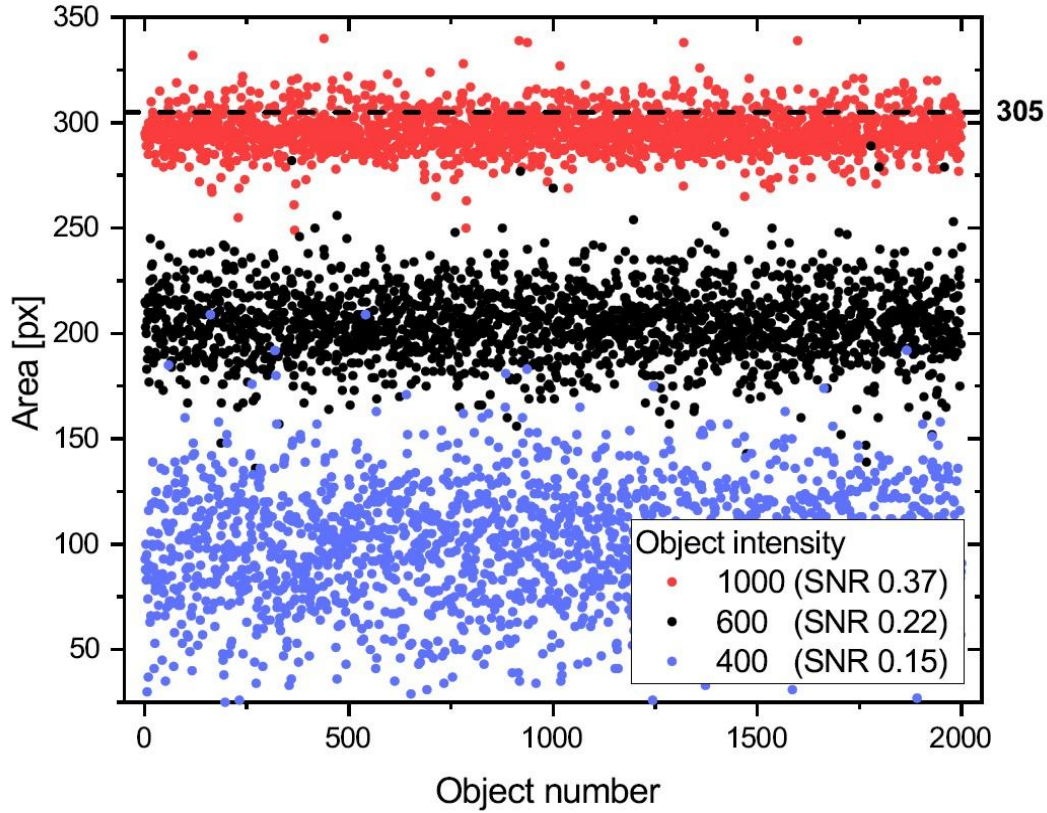

*Fig. 16 Sizes of the detected objects with a radius 10 px and intensities of 1000, 600, and 400. The nominal area of the generated objects is 305 px.*

The next stage aimed to gather information about the number of objects with varying radii and intensities generated against the same background. Unlike the previous test, where each image contained only one type of object, this stage involved multiple different objects generated within a single image. We used the same background images and techniques as in the previous case for this simulation. In each of the 100 generated images, 20 objects were randomly placed. Each object represented a radius of 13, 10, 7, or 5 and had an intensity of 1200, 1100, 1000, 900, or 800. Each object corresponds to a unique radius-intensity pair. In a randomly selected image from those generated, parameters were carefully chosen for the *Background Remover* to ensure that all objects were counted while also minimizing the number of falsely detected objects. Although the objects varied in both intensity and size, we succeeded in selecting the plugin parameters that detected a total of 2001 objects, consisting of 2000 true positives and 1 false positive. It's worth noting that no additional image processing was required, either before or after using the plug-in. A similar image analysis was conducted using the *Background Subtractor* from the Mosaic Suite. Initially, the results appeared promising (see Fig. 17), however, the remaining significant amount of background makes the automatic counting of objects more complicated. This problem can be partially addressed by adjusting the threshold, yet this causes further issues, as depending on the set threshold, some objects tend to either vanish or split into smaller fragments, compromising the accuracy of the results. While the visual outcomes of both plugins appear similar—allowing the user to correctly identify objects in most cases, the images generated by Mosaic *Background Subtractor* are much less suitable for further automatic image processing than those of *Background Remover*.

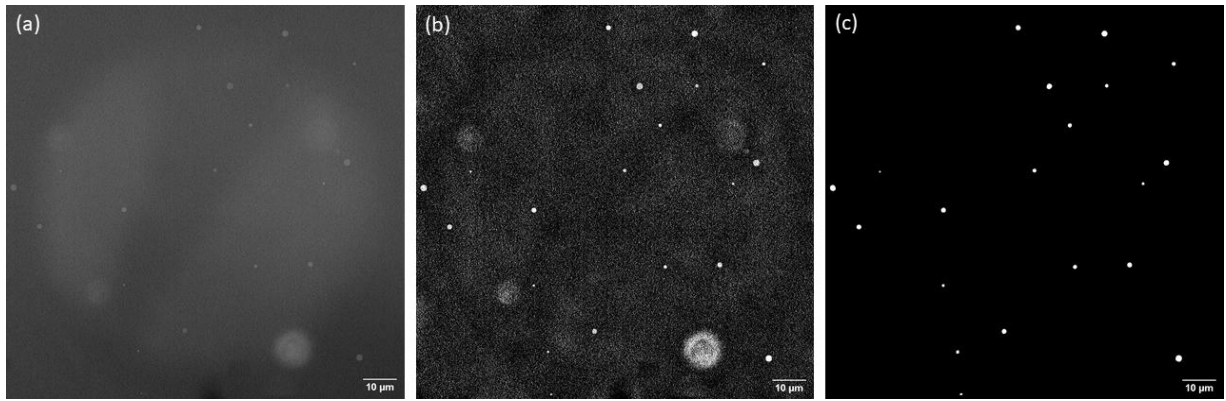

*Fig. 17 Comparison of results obtained using Background Remover and Mosaic Suite's "Background Subtractor": (a) original image, (b) Mosaic Suite's "Background Subtractor" and (c) Background Remover.*

*Test of effectiveness of signal intensity recovering*

The second test aimed to determine the effectiveness of the *Background Remover* plugin in accurately estimating signal intensity (I) on a non-uniform background, which was the main goal of developing the software. An image of the actual background from routine measurements on a lithium fluoride crystal sample was used for the simulation (see Fig. 18). The simulation involved generating 10 objects at randomly selected locations within the image using specified parameters. Measurements were taken for different pairs of object intensity and radius. The radii used were 5 and 10 pixels, and the fixed intensities were 500, 750, 1000, and 2000. The background image pixel values ranged from 227 to 5927, with a mean of 3128 and a median of 3190. This resulted in mean signal-to-noise ratios of 0.16, 0.24, 0.32, and 0.64, respectively. A total of 200 images were generated for each pair of parameters, resulting in 2000 points per pair.

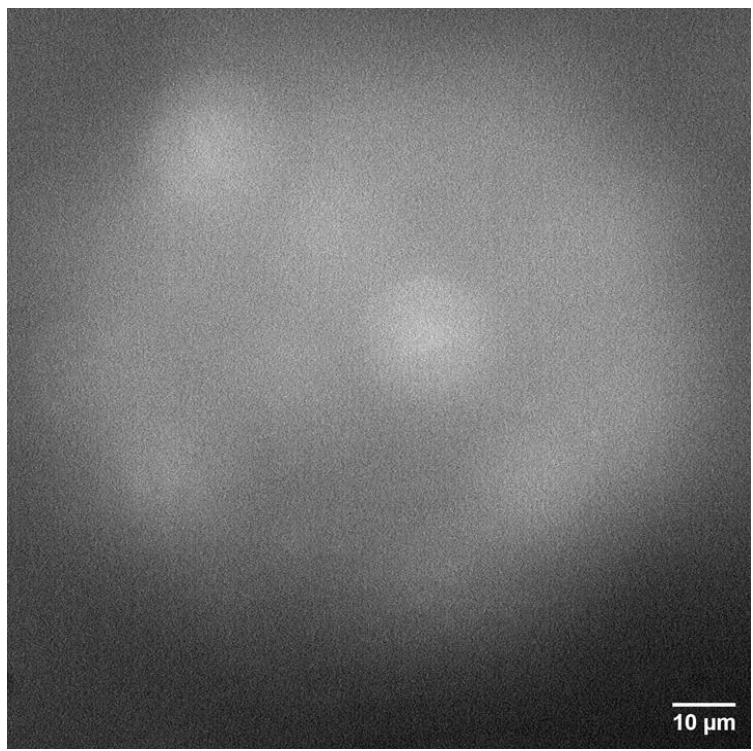

*Fig. 18 Background used in the simulation.*

The manual version of *Background Remover* was utilized to separate the signal from the background, with carefully selected parameters to accurately distinguish the signal from the noise. The *BGR* plugin estimated the local background to obtain the original user-defined object intensity for each of the extracted objects. To do this, the background intensity was calculated as the average of the pixel values in the ring surrounding the point. The ring was created 1 to 4 pixels away from the boundary and was 2 or 3 pixels thick, varying with different radius-intensity pairs. However, it is important to mention that once the parameters were set for a given pair, they were applied to all 200 images with those parameters. *Background Remover* plugin was able to detect all generated objects for all radius-intensity value pairs, except for r=5 px and I=500. In the latter case, only 1851 objects were detected, which is 92.5%.

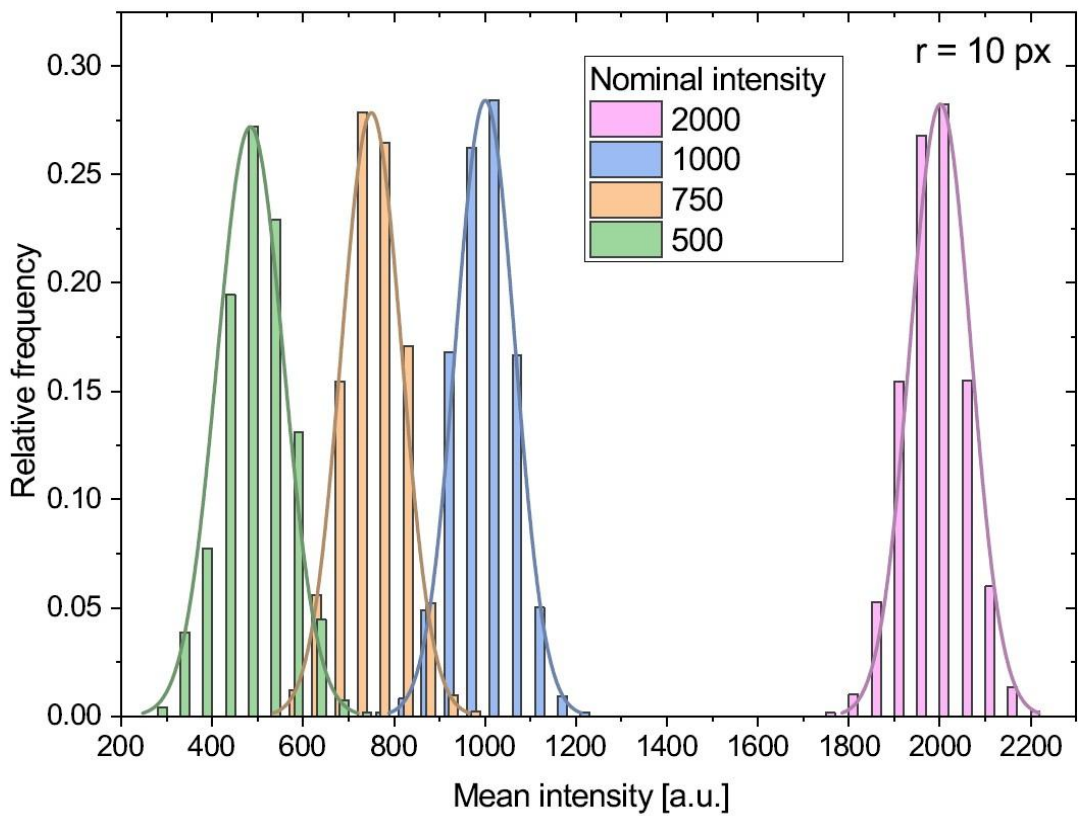

*Fig. 19 Histogram of mean intensity of recovered objects for 10 px radius.*

Using the built-in ImageJ *Analyze Particles* function, the parameters of the recovered objects were analysed, including average object intensity, median object intensity, and area. Figures 19 and 20 show histograms of the mean intensity of recovered objects for both radii. The values in both cases show some dispersion, which is due to two factors: the averaging of the background around the object used by the *Background Remover*, and the necessity to average the background under the objects during their generation. The histograms are well described by normal distributions. The larger width of the distributions for objects with a radius of 5 px results from smaller statistics (305 px area for r = 10 px vs 69 px area for r = 5 px).

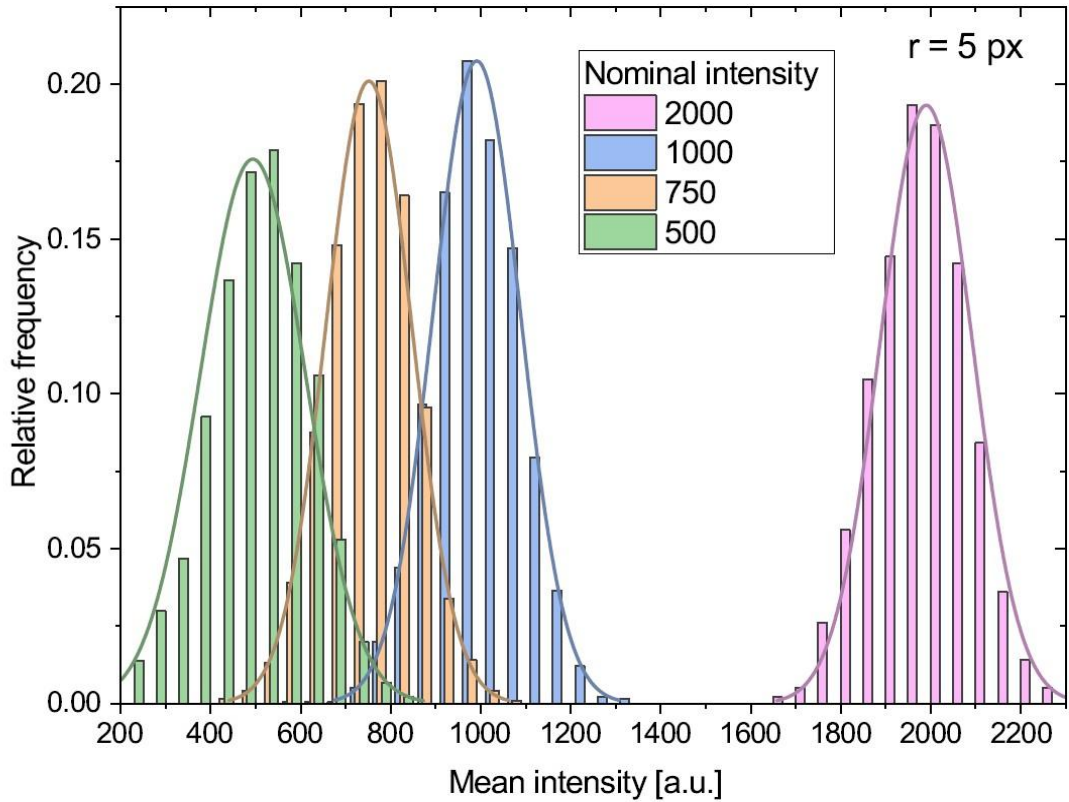

*Fig. 20 Histogram of mean intensity of recovered objects for 5 px radius.*

Table 1 provides a summary of the results obtained. The nominal areas of a 10 px and 5 px circles are 305 px and 69 px, respectively. In all cases, mean and median values consistent with nominal values were obtained, as shown in table 1. The standard deviation is noticeably larger for objects with a diameter of 5 px, which, as mentioned earlier, is due to smaller statistics. When the signal-to-noise ratio decreases, the size of extracted objects also decreases. This occurs due to the necessity of making compromises while adjusting the cutoff parameters. A less strict cutoff criterion preserves more of the detected objects' area but raises the likelihood of false objects detections. Conversely, a stricter criterion reduces background noise more effectively but also removes a percentage of outer pixels of the object.

Table 1 Comparison of the results of recovering information about the intensity of objects. The actual number of objects was always 2000.

| Radius [px] | Nominal intensity [a.u.] | SNR | Mean [a.u.] | Median [a.u.] | Number of detected objects | Area [px] |
|---|---|---|---|---|---|---|
| 10 | 2000 | 0.64 | 2002 ± 67 | 2005 ± 67 | 2000 | 298 ± 3 |
| 10 | 1000 | 0.32 | 1000 ± 65 | 1001 ± 66 | 2000 | 284 ± 11 |
| 10 | 750 | 0.24 | 750 ± 67 | 753 ± 67 | 2000 | 287 ± 12 |
| 10 | 500 | 0.16 | 484 ± 73 | 484 ± 73 | 2000 | 222 ± 15 |
| 5 | 2000 | 0.64 | 1990 ± 102 | 1994 ± 103 | 2000 | 67 ± 2 |
| 5 | 1000 | 0.32 | 991 ± 97 | 1007 ± 101 | 2000 | 66 ± 4 |
| 5 | 750 | 0.24 | 751 ± 96 | 753 ± 96 | 2000 | 50 ± 7 |
| 5 | 500 | 0.16 | 494 ± 117 | 494 ± 116 | 1851 | 24 ± 11 |

**Conclusions**

The *Background Remover* plugin we developed has shown to be an effective tool for extracting small objects from noisy microscopy images. Its local operating mode allows it to function well even with backgrounds that have varying levels of intensity. The plugin not only identifies objects but also preserves their original characteristics, particularly their intensity. Furthermore, it has demonstrated strong performance in estimating the net signal as well. The operating principle based on object identification allows for complete background zeroing, which is particularly useful for subsequent automatic image processing. Although *Background Remover* was developed for a specific type of image, it has proven to be a helpful tool for a wider range of images and we believe that it may be useful also for applications other than FNTD.

A limitation is the user's specific reliance on parameter selection to distinguish objects from the background. An attractive alternative approach for a discrimination stage would be processing each image with a simple discriminative neural network. This could be the next stage of this work when there is a large enough learning set of data.

**Acknowledgments**

This work was supported by the National Science Centre, Poland (grant No 2020/39/B/ST9/00459).

**Appendix 1**

The general expression of a 2D discrete convolution:

$$g(x,y) = \omega * f(x,y) = \sum_{i=-a}^{a} \sum_{j=-b}^{b} \omega(i,j) \cdot f(x-i, y-j)$$

Where:
f(x,y) – original image;
g(x,y) – filtered image/output;
ω - kernel/convolution matrix;
a, b – kernel size;

In our algorithm convolution is performed for each pixel R+1 times. For each time the kernel changes depending on the radius r, which is natural number in the range from 0 to R:

$$g_r(x,y) = \omega_r * f(x,y)$$

The kernel can be described by the following equations:

$$\omega_r(x,y) = \begin{cases} 1 - |r - d|, & d \in (r-1; r+1), \quad d \in (r-1; r+1) \\ 0, & r \in (-\infty; r-1] \cup [r+1; +\infty) \end{cases}$$

$$d = \sqrt{x^2 + y^2}$$

Additionally, it is also necessary to apply normalization:

$$\omega_r^{norm}(x,y) = \frac{\omega_r(x,y)}{\sum_{i=1}^{n} \sum_{j=1}^{n} \omega_r(i,j)}$$

Ultimately we get for a given pixel:

$$Z(R) = g_r^{norm}(x,y)$$

For example, for r=2 and R=3 the kernel takes the form:

$$\begin{pmatrix} 0 & 0 & 0 & 0 & 0 & 0 & 0 \\ 0 & 0.01 & 0.06 & 0.08 & 0.06 & 0.01 & 0 \\ 0 & 0.06 & 0.03 & 0 & 0.03 & 0.06 & 0 \\ 0 & 0.08 & 0 & 0 & 0 & 0.08 & 0 \\ 0 & 0.06 & 0.03 & 0 & 0.03 & 0.06 & 0 \\ 0 & 0.01 & 0.06 & 0.08 & 0.06 & 0.01 & 0 \\ 0 & 0 & 0 & 0 & 0 & 0 & 0 \end{pmatrix}$$

**Appendix 2**

The plugin is actively maintained, with source code, documentation, and example materials available via GitHub and the official project website.

- **GitHub repository**: [https://github.com/kilianna/background-remover/](https://github.com/kilianna/background-remover/)
- **Official documentation and download**: [https://kilianna.github.io/background-remover/](https://kilianna.github.io/background-remover/)
- **Manual movie**: [https://www.youtube.com/watch?v=I8ab4yOq8iU](https://www.youtube.com/watch?v=I8ab4yOq8iU)